\documentclass[prl,aps,twocolumn,showpacs]{revtex4}
\usepackage{graphicx}
\usepackage{amsfonts}

\begin{document}

\title {Derivation  of  hydrodynamics for the  gapless mode  in  the BEC-BCS crossover from the
    exact one-loop effective action}

\author{Da-Shin\ Lee  and Chi-Yong\ Lin}\affiliation{
Department of Physics, National Dong Hwa University, Hua-Lien,
Taiwan 974, R.O.C.  } \author{Ray\ J.\ Rivers}
\affiliation{ Blackett Laboratory, Imperial College\\
London SW7 2BZ, U.K.}

\begin{abstract}
We show that many hydrodynamical properties of the BEC/BCS
crossover in the presence of a Feshbach resonance at $T=0$ can be
derived easily from the derivative expansion of the (exact) fully
renormalized  one-loop effective action. In particular, we
calculate the velocity of sound throughout the BCS and BEC regimes
and derive the generalized superfluid continuity equations for the
composite two -fluid system.
\end{abstract}


\

\maketitle

As the strength of fermionic pairing increases in cold alkali
atoms there is a continuous evolution from the BCS-like behavior
of Cooper pairs to Bose-Einstein Condensation (BEC) of molecules.
This crossover phenomenon becomes controllable
experimentally~\cite{exp} when the atoms interact through a
Feshbach resonance~\cite{holland,timm,griffin, chen,stoof,dieh}.
In both regimes, condensation is a consequence of global $U(1)$
symmetry breaking with a concomitant gapless Goldstone mode (the
phonon).

Although the s-wave scattering length diverges at the crossover,
the speed of sound changes smoothly, as do the hydrodynamical
properties of the condensate. The main result of this {\it letter}
is to show that, at $T=0$, all of these can be determined
straightforwardly from the derivative expansion of the exact,
fully renormalized one-loop effective action, in the spirit of
\cite{dieh}, sidestepping the more complicated derivations based
on a full multi-channel analysis (e.g. see~\cite{griffin}).

Consider a condensate comprising a mixture of fermionic atoms and
molecular bosons. The fermions $\psi_{\sigma} (x)$, with spin
$\sigma = ( \uparrow , \downarrow )$, undergo self-interaction
through an s-wave BCS-type  term. In addition, two fermions can be
bound into a molecular boson $\phi(x)$ through a Feshbach resonance
effect. To exemplify our method, we take the Lagrangian density to
be~\cite{holland,timm} ($U>0$, $g$ fixed)
\begin{eqnarray}
L &=& \sum_{\uparrow , \downarrow} \psi^*_{\sigma} (x)\ \left[ i \
\partial_t + \frac{\nabla^2}{2m} + \mu \right] \ \psi_{\sigma} (x)
\nonumber \\
& +& U \ \psi^*_{\uparrow} (x)\  \psi^*_{\downarrow} (x) \
\psi_{\downarrow} (x) \ \psi_{\uparrow} (x)
\nonumber \\
   &+& \phi^{*}(x) \ \left[ i  \ \partial_t + \frac{\nabla^2}{2M} + 2 \mu -
\nu \right] \ \phi(x) \nonumber \\
&-& g \left[ \phi^{*}(x) \ \psi_{\downarrow} (x) \ \psi_{\uparrow}
(x) + \phi(x) \ \psi^{*}_{\downarrow} (x) \ \psi^{*}_{\uparrow} (x)
\right].
\end{eqnarray}
The mass of the bosons is twice that of the fermions, $M=2m$. The
kinetic energy of the fermions is $\epsilon_{k}=k^2/2m$ and the
kinetic energy of the bosons is $k^2/2M + \nu$, where $\nu$ is the
threshold energy of the Feshbach resonance. To demonstrate the
method we restrict ourselves to a narrow resonance
approximation~\cite{holland,timm} where quantum loop effects from
the molecular boson can be safely ignored~\cite{dieh}, although this
is often an idealization. The effective four-fermion interaction is
determined by the value of $\nu$, which is controllable by an
external magnetic field. On tuning the field, the condensate can be
varied in form from  BCS Cooper pairs to BEC molecules as the
scattering length changes sign.

On introducing the auxiliary field $\Delta (x) = U\psi_{\downarrow}
(x)  \psi_{\uparrow} (x)$, a Hubbard-Stratonovich transformation
leads to an effective Lagrangian density, to which the fermionic
contribution is $\Psi^{\dagger}(x) G^{-1} \Psi(x)$, where $\Psi(x)$
is the Nambu spinor, $\Psi^{\dagger}(x)=( \psi_{\uparrow}^{\dagger},
\psi_{\downarrow} )$, and $G^{-1}$ is the inverse Nambu Green
function
\begin{equation}
 G^{-1} = \left( \begin{array}{cc}
        i \partial_t - \varepsilon         & \tilde{\Delta}(x) \\
                  \tilde{\Delta}^{*}(x) & i \partial_t + \varepsilon
                  \end{array} \right)  \label{greenfun}
                  \end{equation}
with $\varepsilon = - \frac{\nabla^2}{2m} - \mu $.
 The {\it combined} condensate of the theory is
$\tilde{\Delta}(x)$, given in terms of the bifermion and molecular
condensates $\Delta$ and $\phi$ as $\tilde{\Delta}(x) = \Delta(x)
- g \ \phi(x).$ The gapless mode of the theory is encoded in the
phases of $\Delta (x)$ and $\phi (x)$ for which we write $
\Delta(x) = |\Delta(x)|\ e^{i \theta_{\Delta}(x)}, \,\, \phi(x) =
-|\phi(x)| \ e^{i \theta_{\phi}(x)}.$  The amplitude and phase of
$\tilde{\Delta}(x) = |\tilde{\Delta}(x)| \ e^{i
\theta_{\tilde{\Delta}}(x)}$  can be determined from those of
$\Delta (x)$ and $\phi (x)$ by its definition above. We now
perform a $U(1)$ gauge transformation on the fermion field
$\psi_{\sigma}(x)=e^{i \theta_{\tilde{\Delta}}(x)/2} \
\chi_{\sigma}(x)$. Integrating out $\chi_{\sigma}(x)$ leads to a
non-local effective action $ S_{\rm
eff}[\phi,\phi^*,\Delta,\Delta^*]$.

The action possesses a $U(1)$ invariance under the phase change
$\theta_{\phi}\rightarrow\theta_{\phi}+\alpha$ and
$\theta_{\Delta}\rightarrow\theta_{\Delta}+\alpha$ (or
$\theta_{\tilde{\Delta}}\rightarrow\theta_{\tilde{\Delta}}+\alpha$).
This symmetry is spontaneously broken when  $\Delta$ and $\phi$
respectively acquire the non-vanishing constant values $\Delta_0$
and $\phi_0$ determined by the  gap equations obtained from
extremizing the effective action. In these, $|\tilde{\Delta}_{0}|
= |\Delta_{0}| + |-g\phi_0|$ satisfies
\begin{equation}
\frac{1}{U_{\rm eff}} = \int \frac{ d^{3} {\bf {p}}}{  (2 \pi)^3} \,
\frac{1}{ 2E_{p}}, \label{d-condensate}
\end{equation}
with $ U_{\rm eff} = U + g^2/ (\nu-2\mu) $ and $E_{p}=(
\varepsilon_{p}^2 + |\tilde{\Delta}_{0}|^2 )^{1/2} $, and
\begin{equation}
 |\phi_{0}|= \frac{g}{\nu - 2\mu} \ \frac{ | \tilde{\Delta}_{0}
| }{ U_{ {\rm eff} } }\,  \label{m-condensate}.
\end{equation}

Since the microscopic theory under consideration is Galilean
invariant, any effective theory derived from it must respect this
symmetry~\cite{pit}. Consider the dynamics of the phonon carried in
the angular variables. To preserve Galilean invariance at each step,
the variations in the condensate magnitudes can be written as $
|\Delta| = |\Delta_{0}| + \delta |\Delta|, \,\,\,\, |\phi| =
|\phi_{0}| + \delta |\phi|$~\cite{aitchison,stone,palo}. We assume
that terms in $\delta |\tilde\Delta|$, $\delta |\phi|$, and
$(\theta_{\Delta}-\theta_{\phi})^{2}$ are of the same order in their
defining equation, although $\theta_{\Delta}$ and $\theta_{\phi}$
are large variables. We now use the fact that $e^{-i\sigma_3
\theta_{\tilde{\Delta}}(x)/2}G^{-1}e^{i\sigma_3
\theta_{\tilde{\Delta}}(x)/2} = G_{0}^{-1} - \Sigma$, where
\begin{eqnarray}
\Sigma &=& ( - i \nabla^{2} \theta_{\tilde{\Delta}} /4m + (\nabla
\theta_{ \tilde{\Delta}    } ) (-i \nabla)/2m) {\rm I}\\
\nonumber
   &+& ( \dot{\theta}_{\tilde{\Delta}}/2 + (\nabla \theta_{
\tilde{\Delta} })^2/ 8m ) \sigma_{3} - \delta |\tilde{\Delta}| \
\sigma_{1},
\end{eqnarray}
 with $\delta |\tilde{\Delta}|$ defined by
$ |\tilde{\Delta}| = |\tilde{\Delta_{0}}| + \delta
 |\tilde{\Delta}|$~\cite{aitchison}. $
G_{0}^{-1}$ is the free inverse Nambu Green function with the same
form as $ G^{-1}$ in  Eq.({\ref{greenfun}), except that
$\tilde{\Delta}(x)$ is replaced by $|\tilde{\Delta}_{0}|$. $
S_{\rm eff}[\phi,\phi^*,\Delta,\Delta^*]$ then permits the
derivative expansion
\begin{eqnarray}
S_{ {\rm eff} } &=& -i {\rm Tr} {\rm Ln} (G_{0}^{-1}) + i {\rm Tr}
\sum_{n=1}^{\infty} \frac{(G_{0} \Sigma)^{n}}{n} -\int d^4 x
\frac{|\Delta |^2}{U} \nonumber \\
&+& \int d^{4} x \phi^{*}(x) \left( i\partial_{t}  +
\frac{\nabla^2}{2M} + 2\mu - \nu \right) \phi(x), \label{effaction}
\end{eqnarray}
the first two terms of which are no more than the expansion of $-i
{\rm Tr} {\rm Ln} (G^{-1})$. For our purposes it is sufficient to
truncate the sum in $n$ at second order, to yield an action
$S^{(2)}_{{\rm eff}}[\delta|\phi |,\delta|\tilde{\Delta}|,
\theta_{\phi},\theta_{\tilde{\Delta}}]$ after eliminating $\Delta$
in favor of $\tilde{\Delta}$, and $\theta_{\Delta}$ in terms of
$\theta_{\tilde{\Delta}}$, on using $(\theta_{\tilde{\Delta}} -
\theta_{\phi})\approx(|\Delta_{0}|/|\tilde{\Delta}|)(\theta_{\Delta}
- \theta_{\phi})$. Diagrammatically, this amounts to taking
account of  fermionic one-loop effects in the effective action}.

 The  action of the gapless phonon mode is that part of the quadratic
contribution to $S^{(2)}_{\rm eff}$ in which the derivatives of
$\delta|\phi |$ and $\delta|\tilde{\Delta}|$ are omitted.
 After straightforward manipulations this takes the form
\begin{eqnarray}
 S_{{\rm phonon}} &=& \int d^4x\bigg\{ \frac{N}{4}\dot{\theta}_{\tilde{\Delta}}^2
- \frac{1}{8m} \rho^F_0(\nabla \theta_{\tilde{\Delta}} )^2
 -\frac{1}{8m}\rho^B_0 (\nabla
\theta_{\phi} )^2 \nonumber \\
&-& \frac{1}{2}\Omega^2(\theta_{\tilde{\Delta}}-\theta_{\phi})^2 - 2
|\phi_{0}| \delta |\phi| \ \dot{\theta}_{\phi}
-\alpha\,\dot{\theta}_{\tilde{\Delta}}\delta|\tilde{\Delta}|
\nonumber \\
&+& (2\mu - \nu)\frac{U_{\rm eff}}{U}(\delta |\phi|)^2 +
\frac{2g}{U}\delta|\tilde{\Delta}|\delta |\phi| \nonumber \\
&-& \frac{1}{2}\bigg(\frac{2}{U} -\beta
  \bigg)(\delta|\tilde{\Delta}|)^2
\bigg\}. \label{Squad}
\end{eqnarray}

This lends itself to a very simple mechanical picture of a coupled
'wheel' and 'axle'. The radius of the wheel is $|\tilde{\Delta}|$
and that of the axle is $|-g\,\phi |$, measured from the gap
values $|\tilde{\Delta}_0|$ and $|-g\,\phi_0 |$ with angles of
displacement $\theta_{\tilde{\Delta}}$, $\theta_{\phi}$,
respectively. There is slippage between the wheel and axle with an
elastic restoring force $\Omega^2=(2 g/ U)
|\phi_{0}||\tilde{\Delta}_{0}|$.

The fermion number density at $n=1$  is
\begin{equation}
 \rho_0 = \rho^F_0 +
\rho^B_0 , \label{f-density}
\end{equation}
  where $
 \rho^F_0 = \int d^3 {\bf p} / (2\pi)^3 \ \left[ 1 -
\varepsilon_{p}/E_{p}
  \right] $
is the explicit fermion density, and $\rho^B_0 = 2|\phi_{0}|^2$ is
due to molecules (two fermions per molecule). The other
coefficients are straightforwardly derived as $ N = \int d^{3}
{\bf p}/ (2\pi)^3 ( |\tilde{\Delta}_{0}|^2 / 2 E_{p}^3), \alpha =
\int d^3 {\bf p} / (2\pi)^3 (|\tilde{\Delta}_0|\varepsilon_{p}/
2E_{p}^3),$ and $
 \beta = \int d^3 {\bf p} / (2\pi)^3
(\varepsilon_{p}^2/ E_{p}^3) \, .$

A further straightforward eigenvalue calculation gives the long
wavelength dispersion relation for the phonon as $\omega^2 =
v^2{\vec k}^2 + O(k^4)$, where
 \begin{equation}
 v^2 = \frac{\rho_0/2m}{N + A/B }
 \label{v}
 \end{equation}
 and
 \begin{eqnarray}
A &=& \alpha^2(2\mu -\nu)\frac{U_{\rm eff}}{U}
  -2\alpha |\phi_0|\frac{2g}{U} - 2 |\phi_0|^2\bigg(\frac{2}{U} -\beta
  \bigg),
  \nonumber \\
 B &=& \bigg(\frac{g}{U}\bigg)^2 + \frac{1}{2}\bigg(\frac{2}{U} -\beta
  \bigg)(2\mu -\nu)\frac{U_{\rm eff}}{U}.\label{numdom}
  \end{eqnarray}
Note that $v$ is independent of the slippage strength $\Omega\neq
0$.

\begin{figure}
\centering
\includegraphics[width=\columnwidth]{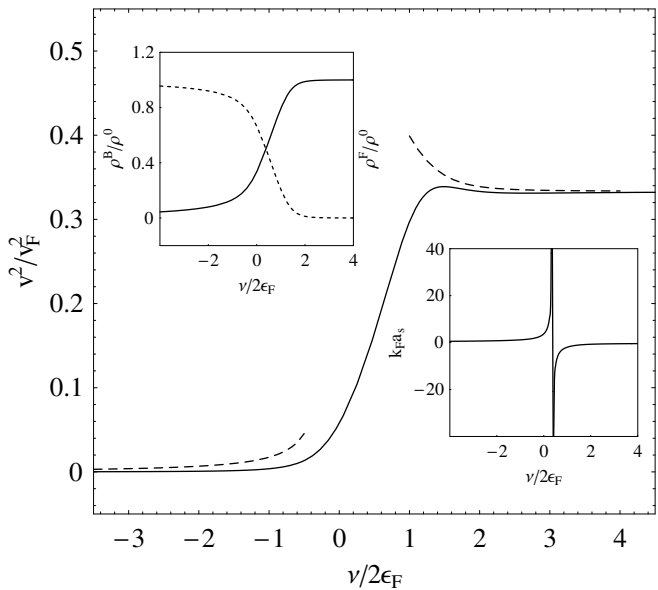}
 \caption{The behavior of the sound velocity $ v $ throughout BEC and BCS regimes as a function of the
 threshold energy $\nu$. We choose $U= 7.54 \, \epsilon_F/k^3_F$
and $ g=4.62 \, \epsilon_F/ k^{3/2}_F$ as an example~\cite{chen1}.
The dotted line is obtained
 with the approximate solution of the sound velocity, Eq.(\ref{vBEC})
 for the BEC regime and Eq.(\ref{vBCS})
 for the BCS regime.  The lower inset shows how the scattering length $a_s$ varies  from BEC
to BCS as the threshold energy increases, while the upper inset
reveals the evolution of  the fermion density $\rho^F_0$ (solid
line) and the molecule density
 $\rho^B_0$ (dotted line).} \label{fig2}
\end{figure}

In fact, $\beta$ is UV-singular, as are $U_{eff}, U$ and $g$ from
the gap equations (\ref{d-condensate}) and (\ref{m-condensate}).
Renormalization is implemented by defining the renormalized
coupling $\bar{U}_{\rm eff}$ in terms of the s-wave scattering
length $a_s$ as
\begin{eqnarray}
-\frac{N_0}{k_Fa_s} &=& \frac{1}{\bar{U}_{\rm eff}} =
\frac{1}{U_{\rm eff}} - \int^{\Lambda} \frac{ d^{3} {\bf p} }{ (2
\pi)^3 } \
\frac{1}{2\epsilon_{p}} \nonumber \\
&=& \int^{\Lambda} \frac{ d^{3} {\bf p} }{ (2 \pi)^3 } \
\bigg[\frac{1}{2E_{p}}-\frac{1}{2\epsilon_{p}}\bigg], \label{aS}
\end{eqnarray}
where $k_F$ is the Fermi momentum and $\Lambda$ is a UV cutoff.
UV-finite renormalized couplings $\bar{U}$ and $\bar{g}$ are defined
similarly in the limit $\Lambda\rightarrow\infty$;
\begin{equation}
\frac{1}{\bar{U}}=\frac{1}{U} - \int^{\Lambda} \frac{ d^{3} {\bf p}
}{ (2 \pi)^3 } \ \frac{1}{2\epsilon_{p}} \, ,\,\,\,\,
\frac{1}{\bar{g}}=\frac{1}{g} - \frac{U}{g}\int^{\Lambda} \frac{
d^{3} {\bf p} }{ (2 \pi)^3 } \ \frac{1}{2\epsilon_{p}}. \label{Ubar}
\end{equation}
In turn, we now define a renormalized threshold energy $\bar{\nu}$
through
\begin{equation}
 \bar{U}_{\rm eff} = \bar{U} + \frac{\bar{g}^2}{\bar{\nu}-2\mu}
\, .
\end{equation}
The outcome of this renormalization is that  $S_{{\rm phonon}}$ of
(\ref{Squad}) and the gap equation for the condensate
(\ref{m-condensate}) are rendered UV finite term by term by
replacing the unrenormalized $U, U_{\rm eff}, g, \beta, \nu$ by
their renormalized counterparts. Henceforth we drop the overbars
for simplicity, and understand all quantities in (\ref{v}) as
renormalized. Necessarily, the renormalization prescription we
propose above makes not only the gap equations free of
UV-divergence, but also the sound velocity. Thus, combining the
renormalized equations (\ref{m-condensate}) and (\ref{aS}) with
the number equation (\ref{f-density}) allows us to study the
behavior of the condensate numerically when the threshold energy
varies so that the
 scattering length goes from  $a_s\rightarrow 0^+$, the deep BEC regime, to
 $a_s\rightarrow 0^-$, that of  deep BCS, through a BCS/BEC crossover. This is shown
 in the inset in
Fig.(\ref{fig2})~\cite{stoof, chen1}. Solving first for the
chemical potential $\mu$ as a function of the threshold energy
$\nu$, the sound velocity (\ref{v}) can be computed, as depicted
in the main Fig.(\ref{fig2}), in agreement with \cite{stoof2}.

To understand the numerical behavior shown in Fig.(\ref{fig2}) we
evaluate $v^2$ analytically in both deep BCS and BEC regimes. In the
deep BCS regime, there is considerable simplification as all
momentum integrals are dominated by the $ {\bf k}$ modes near the
Fermi surface. In this limit,  $|\phi_0| \approx 0$, where few
molecules are present,  and $\alpha \approx 0$, the ignorable
amplitude-phase coupling due to the particle-hole symmetry near the
Fermi surface. We obtain $A/B\ll N$, leading to
 \begin{equation}
 v^2\simeq \frac{\rho_0}{2mN} .
 \label{vBCS}
 \end{equation}
 With the total fermion density $\rho_0 = k_F^3/3\pi^2$ and $N
\approx N_0= mk_F/2\pi^2$, the fermion density of states at the
Fermi surface, the sound velocity is given by $ v^2 \simeq v^2_F/3
$. This is the result we obtain from conventional BCS theory with
no Feshbach resonance (e.g. see \cite{aitchison}), for which $g=0$
and $|\phi_0| = 0$.

On the other hand, in the deep BEC limit ${U}\ll {g}^2/|{\nu} -
2\mu|$, which gives rise to
 $|{\nu} - 2\mu|
\approx g^2 N_0/ k_F a_s $, as obtained from $ U_{\rm eff}$. To
maintain this relation for small $ a_s$,  $ U$  cannot be too large.
From Eqs. (\ref{numdom}), the sound velocity $ v^2 $ can be
approximated as
\begin{equation}
v^2\simeq \frac{1}{8m}\frac{|\tilde\Delta_0|^2}{|\mu|} \,
\label{vBEC}
\end{equation}
in this deep BEC regime as a result of $ A/B \gg N$. We find that
$|\phi_0|$ increases as ${\nu}$ goes increasingly negative, while
the combined condensate $|\tilde{\Delta}_0|$ decreases to keep the
number of fermions fixed.
 As a result $ |\tilde{\Delta}_0|^2 \simeq g^2|\phi_0|^2\simeq (g^2/ 2)
\rho_0$ for largely  negative $\nu$ (i.e. $a_s \rightarrow 0^+$)
when all fermions are in the form of molecules as seen in the
inset in Fig.(1).
 Thus, the behavior of $ v^2 $, which is found to approach
zero at small $a_s$, is now determined by how the chemical
potential $\mu$  increases negatively as $ a_s$ deceases.

In the central region, the behavior is smooth across the 'unitarity
limit' at $|a_s|\rightarrow\infty$ when $\nu=2\mu$. In this case,
the gap and number equations are reduced to involving only one
parameter $ g $ in the one-channel formulation of Ref.\cite{ho}. If
$ g$ is large, 'universal behavior' is found where $ v^2 \approx 0.2
v^2_F$. Nevertheless,  the renormalization of the molecular boson is
expected to contribute sizable corrections to the sound velocity
obtained above in such a strongly  coupled regime~\cite{astrak}.

More generally, the BEC/BCS system permits a hydrodynamic
interpretation as a two-component superfluid. The crucial
ingredient is the Galilean invariance of the derivative expansion.
It is not difficult to rederive the relevant angular parts of
$S^{(2)}_{\rm eff}$ from $S_{\rm phonon}$. We restore Galilean
invariance by the substitutions
\begin{equation}
\dot{\theta}\rightarrow \dot{\theta} + \frac{(\nabla \theta
)^2}{4m},\,\,\,\,\,\frac{(\nabla \theta )^2}{4m}\rightarrow
\dot{\theta} + \frac{(\nabla \theta )^2}{4m}
\end{equation}
for both phase angles.

Taking the variation of $S^{(2)}_{\rm eff}$ with respect to the
condensate phase $\theta_{\tilde{\Delta}}$ leads to a single
mean-field equation of motion which can be rewritten in terms of
the explicit fermion number density $\rho_{F}$:
 \begin{equation}
 \frac{\partial}{\partial t}\rho_{F} + {\bf\nabla}
 \cdot{\bf j}_{F} - 2{\Omega}^2 (\theta_{\tilde{\Delta}}
 -\theta_{\phi}) = 0,\label{hydroF}
 \end{equation}
 where
 \begin{equation}
 \rho_{F}= \rho^0_F - N_0\bigg({\dot\theta_{\tilde{\Delta}}}+\frac{(\nabla
 \theta_{\tilde{\Delta}}
)^2}{4m}\bigg) + 2\alpha\,\delta {\tilde\Delta}
\end{equation}
and ${\bf j}_{F}= \rho_{F} {\bf\nabla}\theta_{\tilde{\Delta}}/2m$.
There is a similar equation for the fermion number density due to
molecules:
 \begin{equation}
 \frac{\partial}{\partial t}\rho_{B} + {\bf\nabla}
 \cdot{\bf j}_{B} + 2{\Omega}^2 (\theta_{\tilde{\Delta}}
 -\theta_{\phi}) = 0,\label{hydroB}
 \end{equation}
 where (to lowest order)
 \begin{equation}
 \rho_{B} =\rho^0_B + 4|\phi_0|\delta\phi = 2|\phi |^2,
 \end{equation}
 and
 ${\bf j}_{B} = \rho_{B}
{\bf\nabla}\theta_{\phi}/4m$. Putting these together gives the
continuity equation for total fermion number
\begin{equation}
 \frac{\partial}{\partial t}\rho + {\bf\nabla}
 \cdot{\bf j} = 0,\label{hydro}
 \end{equation}
 describing a coupled two-component superfluid, where $\rho = \rho_F + \rho_B$
 and ${\bf j} = {\bf j}_F + {\bf j}_B$.
The behaviour of the mean explicit and molecular fermion number
densities $\rho_F^0$ and $\rho_B^0$ as $\nu$ varies is given in
the inset of Fig.1.

Our earlier definitions give $\Omega^2(\theta_{\tilde{\Delta}} -
\theta_{\phi}) \propto |\Delta_{0}||\phi_0 |(\theta_{\Delta} -
\theta_{\phi})$, showing that the coupling between the superfluid
components is due to the difference in the phase of the fermionic
pairs $\Delta$ and the molecular field $\phi$. It vanishes in both
the deep BCS and BEC regimes, when $|\phi_0 |$ and $|\Delta_{0}|$
tend to zero respectively. Further, with $\rho_F\gg \rho_B$
 and $|{\bf j}_F|\gg|{\bf j}_B|$ in the
deep BCS regime and $\rho_F\ll \rho_B$
 and $|{\bf j}_F|\ll|{\bf j}_B|$ in the deep BEC regime, the system is
 described by a single fluid in each case.
 Away from these extremes the situation gets more complicated,
 with the coupling strongest in the transition regime,
 but still tractable for vortices with phase coupling $\theta_{\Delta}
 =\theta_{\phi}$, whose properties will be pursued elsewhere.

Finally, we briefly consider Gross-Pitaevskii (GP) equations in
the BEC regime. (For the BCS regime the results of
\cite{aitchison} can be simply extended to $g\neq 0$). The
relationship of superfluid equations to GP equations is well
established, in principle. To each ($\rho, \bf j$) pair there is
allocated a complex GP (or non-linear Shroedinger) field $\Psi$.
In our case the Gross-Pitaevskii fields underlying (\ref{hydroF})
and (\ref{hydroB}) are $\Psi_F = \sqrt{\rho_F}
e^{i\theta_{\tilde{\Delta}}}/\sqrt{2}$ and $\Psi_B = \phi
=\sqrt{\rho_B} e^{i\theta_{\phi}}/\sqrt{2}$. Although, in general,
$\Psi_F$ has the phase of the combined condensate, but the
magnitude due to the explicit fermion density only, it happens
that, in the BEC regime,
\begin{equation}
|\Psi_F|^2=
\frac{1}{32\pi}\frac{(2m|\mu|)^{3/2}}{|\mu|^2}|\tilde{\Delta}|^2.
\label{deltapsi}
\end{equation}
 That is, now $\Psi_F\propto\tilde{\Delta}$, linking its phase to the
condensate density.

 Furthermore, $S_{\rm phonon}$ is all that is
needed to extract the coupling constant of two-body interactions
between the condensate. The form of this two-body interaction is
\begin{equation}
 \lambda_{\tilde{\Delta}}(\tilde{\Delta}^{\dagger}\tilde{\Delta})^2
 = \lambda_{\tilde{\Delta}}(|\tilde{\Delta}_0|^4 +
 4|\tilde{\Delta}_0|^2(\delta |\tilde{\Delta}|)^2 +...)
\end{equation}
 From the second term we can read off $\lambda_{\tilde{\Delta}}$ directly from
$S_{\rm phonon}$ as $\lambda_{\tilde{\Delta}}= (2/U -
\beta)/8|\tilde{\Delta}_0|^2$. Again in the BEC regime, we find
\begin{equation}
\lambda_{\tilde{\Delta}}\simeq
-\frac{1}{256\pi}\frac{(2m|\mu|)^{3/2}}{|\mu|^3}. \label{condint}
\end{equation}

 Let us rewrite
$\lambda_{\tilde{\Delta}}(\tilde{\Delta}^{\dagger}\tilde{\Delta})^2$
as a GP self-interaction $\lambda |\Psi_F|^4$, using the
renormalization of (\ref{deltapsi}). On using (\ref{condint}) we
find a weakly repulsive interaction $\lambda =
-2\pi|a_{\tilde{\Delta}}|/M$, where $a_{\tilde{\Delta}} = 2a_s$ and
$M=2m$, as follows from the strong-coupling Bogoliubov-de Gennes
equations \cite{strinati}.

The extension of our approach to non-zero temperature is
straightforward, in principle, and will be considered elsewhere.
It has yet to be seen whether, in general, the effect of Landau
damping can be interpreted as a normal fluid component in addition
to the coupled superfluids of (\ref{hydro}), as happens for pure
BCS theory \cite{metikas}.

 DSL would like to thank The Royal Society for support and
  RR would like to thank the ESF COSLAB programme. We thank Georgios Metikas for
 helpful conversations. This work of DSL and CYL was supported in
 part by the National Science Council,
Taiwan, R.O.C..

\end{document}